\documentclass[useAMS,usenatbib]{mn2e}

\usepackage{graphicx}
\bibpunct{(}{)}{;}{a}{}{,}
\usepackage{txfonts}

\newcommand{\ion}[2]{#1\,{\sc #2}}

\title[Magnetic field in SFXT binaries]{
A search for the presence of magnetic fields in the two Supergiant Fast X-ray Transients
IGR\,J08408$-$4503 and IGR\,J11215$-$5952
}

\author[Hubrig et al.\ 2017]{
S.~Hubrig$^1$\thanks{E-mail: shubrig@aip.de},
L.~Sidoli$^2$,
K.~Postnov$^3$,
M.~Sch\"oller$^4$,
A.~F.~Kholtygin$^5$,
S.~P.~J\"arvinen$^1$,
\and P.~Steinbrunner$^6$\\
$^1$ Leibniz-Institut f\"ur Astrophysik Potsdam (AIP), An der Sternwarte~16, 14482~Potsdam, Germany\\
$^2$ INAF, Istituto di Astrofisica Spaziale e Fisica Cosmica, Via E. Bassini 15, 20133 Milano, Italy\\
$^3$ Sternberg Astronomical Institute, Moscow M.V. Lomonosov State University, 119234 Moscow, Russia\\
$^4$ European Southern Observatory, Karl-Schwarzschild-Str.~2, 85748~Garching, Germany\\
$^5$ Saint-Petersburg State University, Universitetskij pr.~28, Saint-Petersburg 198504, Russia\\
$^6$ Freie Universit\"at Berlin, Kaiserswerther Str.~16-18, 14195~Berlin, Germany
}

\begin{document}

\date{Accepted Received; in original form}

\pagerange{\pageref{firstpage}--\pageref{lastpage}} \pubyear{2015}

\maketitle

\label{firstpage}

\begin{abstract}
A significant fraction of high-mass X-ray binaries are supergiant fast X-ray transients (SFXTs). The prime model 
for the physics governing their X-ray behaviour suggests
that the winds of donor OB supergiants are magnetized. To investigate if magnetic fields are indeed present in
the optical counterparts of such systems, we acquired low-resolution spectropolarimetric 
observations of the two optically brightest SFXTs, IGR\,J08408$-$4503 and IGR\,J11215$-$5952 with the ESO FORS\,2 instrument 
during two different observing runs. No field detection at a significance level of 3$\sigma$ was achieved 
for IGR\,J08408$-$4503. For IGR\,J11215$-$5952, we obtain 3.2$\sigma$ and 3.8$\sigma$ detections 
($\left< B_{\rm z}\right>_{\rm hydr}=-978\pm308$\,G and $\left< B_{\rm z}\right>_{\rm hydr}=416\pm110$\,G) on 
two different nights in 2016. These results indicate that the 
model involving the interaction of a magnetized stellar wind with the neutron star magnetosphere can
indeed be considered to characterize the behaviour of SFXTs.
We detected long-term spectral variability in IGR\,J11215$-$5952, while for IGR\,J08408$-$4503 we find 
an indication of the presence of short-term variability on a time scale of minutes.
\end{abstract}


\section{Introduction}

Among the bright X-ray sources in the sky, a significant number contain a compact 
object (either a neutron star or a
black hole) accreting from the wind of a companion star with a mass above $10\,M_\odot$ \citep{Liu2006}.
Such systems are 
called high-mass X-ray binaries (HMXBs). They are young (several
dozen million years old) and can be formed when one of the initial binary members 
loses a significant part of its mass through stellar wind or mass transfer before a first
supernova explosion occurs \citep{Heuvel1972}. 
Studies of different types of HMXBs are of special interest 
to obtain reliable predictions about the populations of relativistic binaries such as 
double degenerate binaries, which are considered to be gravitational wave progenitors.
Indeed, just a few weeks ago, scientists have directly detected gravitational waves 
from the spectacular collision of two neutron stars (e.g., \citealt{Cowperthwaite2017}). 
Furthermore, high-mass X-ray binaries are fundamental for studying stellar evolution,
nucleosynthesis, structure and evolution of galaxies, and accretion processes. 

Supergiant Fast X-ray Transients (SFXTs) are a subclass of HMXBs associated with early-type
supergiant companions, and characterized by sporadic, short and bright X–ray flares
reaching peak luminosities of 10$^{36}$--10$^{37}$\,erg\,s$^{-1}$ and typical energies released in 
bright flares of about 10$^{38}$--10$^{40}$\,erg (see the review of \citealt{Sidoli2017} for more details). 
Their X-ray spectra in outburst 
are very similar to accreting pulsars in HMXBs. In fact, half of them have measured neutron
star spin periods similar to those observed from persistent HMXBs (\citealt{Shakura2015,Martinez2017}).

\begin{table}
\caption{
Logbook of the FORS\,2 polarimetric observations of IGR\,J08408$-$4503 and IGR\,J11215$-$5952,
including the modified Julian date of mid-exposure,
followed by the achieved signal-to-noise ratio in the Stokes~$I$ spectra around 5200\,\AA{},
and the measurements of the mean longitudinal magnetic field using the Monte Carlo bootstrapping test,
for the hydrogen lines and for all lines.
In the last columns,
we present the results of our measurements using the null spectra for the set
of all lines, and the orbital phase (see text).
All quoted errors are 1$\sigma$ uncertainties. 
}
\label{tab:log_meas}
\centering
\begin{tabular}{lrr@{$\pm$}rr@{$\pm$}rr@{$\pm$}rr}
\hline
\hline
\multicolumn{1}{c}{MJD} &
\multicolumn{1}{c}{SNR} &
\multicolumn{2}{c}{$\left< B_{\rm z}\right>_{\rm hydr}$} &
\multicolumn{2}{c}{$\left< B_{\rm z}\right>_{\rm all}$} &
\multicolumn{2}{c}{$\left< B_{\rm z}\right>_{\rm N}$} &
\multicolumn{1}{c}{$\varphi_{\rm orb}$}\\
\multicolumn{1}{c}{57000+} &
\multicolumn{1}{c}{$\lambda$5200} &
\multicolumn{2}{c}{[G]} &
\multicolumn{2}{c}{[G]}  &
\multicolumn{2}{c}{[G]} &
\\
\hline
\multicolumn{9}{c}{IGR\,J08408$-$4503} \\
\hline
    727.3424& 2730 &    21  & 132 &   162  & 114 &    87 & 135& 0.080\\
    736.2242& 3391 & $-$184 &  97 & $-$29  &  59 & $-$37 &  62& 0.011\\
    745.2449& 1844 &    269 & 158 &   131  & 100 & $-$93 &  82& 0.956\\
    747.2578& 3401 & $-$141 &  94 &    18  &  67 & $-$39 &  63& 0.167\\
\hline
\multicolumn{9}{c}{IGR\,J11215$-$5952} \\
\hline
    522.0581& 1012 &  $-$978& 308 & $-$646 & 317 &     0 &356 & 0.958\\ 
    528.0357& 1926 &     406& 156 &    222 &  93 & $-$44 & 93 & 0.995\\
    738.2863& 2844 &     416& 110 &    208 &  62 &    31 & 67 & 0.272\\ 
    764.2995& 1957 &   $-$96& 145 &    116 &  73 &$-$143 & 88 & 0.430\\ 
    793.3257& 2849 &     241& 113 &    133 &  55 &    67 & 48 & 0.606\\
\hline
\end{tabular}
\end{table}

The physical mechanism driving their transient behavior,
probably related to the accretion of matter from the supergiant wind by the compact object,
has been discussed by several authors and is still
a matter of debate. The prime model for the existence of SFXTs
invokes their different wind properties and magnetic field strengths,
which lead to distinctive accretion regimes
(\citealt{Shakura2012,Shakura2014,Shakura2017}).
The SFXTs' behaviour can be explained 
by sporadic capture of magnetized stellar wind. 
The effect of the magnetic field carried by the stellar wind is twofold: first, it may
trigger rapid mass entry to the magnetosphere via magnetic reconnection in the 
magnetopause (a phenomenon that is well known in the dayside of the Earth 
magnetosphere), and
secondly, the magnetized parts of the wind (magnetized clumps with a tangent 
magnetic field) have a lower velocity than the non-magnetized parts or the parts carrying the
radial field \citep{Shakura2014}.
The model predicts that a magnetized clump 
of stellar wind with a magnetic field strength of a few tens of Gauss triggers  
sporadic reconnection, allows accretion, and results in an X-ray flare.
Since, typically, the orbital separation between the neutron star and the supergiant
is a few $R_\ast$ of the latter,
the expected required magnetic field on the stellar surface should be of the order of 
$100-1000$\,G.

To investigate the magnetic nature of the optical counterparts in these systems,
we recently observed the two optically brightest SFXTs, IGR\,J08408$-$4503 ($m_{V}=7.6$) and 
IGR\,J11215$-$5952 ($m_{V}=10.0$), using the FOcal Reducer low dispersion 
Spectrograph (FORS\,2; \citealt{Appenzeller1998}) mounted on the 8\,m Antu telescope of 
the VLT.
The optical counterpart of the system IGR\,J08408$-$4503 is the O-type
supergiant star HD\,74194 with spectral classification O8.6\,Ib-II(f)p \citep{Sota2014}.
The orbital solution ($P_{\rm orb}=9.5436\pm0.002$\,d, $e=0.63\pm0.03$) was for the first time determined by 
\citet{Gamen2015}. This work also indicated that the equivalent width of the H$\alpha$ line
is not modulated entirely with the orbital period, but seemed to vary with a superorbital period
($P=285\pm10$\,d) nearly 30 times longer than the orbital one. This orbital solution also supported 
the existence of a correlation between high-energy outbursts and periastron passages.

The astrophysical parameters of the optical counterpart of the system IGR\,J11215$-$5952, 
the B0.5\,Ia supergiant HD\,306414, were recently studied by \citet{Lorenzo2014},
who suggested that this supergiant of about 30\,$M_\odot$ has just completed H core burning, showing only 
slight enhancement of He and N. No clear orbital variability, neither in radial velocities nor in the 
photometric light curve, was detected. 
X-ray outbursts occurring every 164.6\,d  (\citealt{Sidoli2007,Romano2009}) 
strongly suggest that this recurrence timescale is the orbital period. 
In this case, the outbursts are thought to be produced near the periastron passage, although the 
real orbital phase where they occur is unknown.

High-resolution
optical  spectroscopy with the FEROS instrument mounted on the ESO/MPG\,2.2\,m telescope on La Silla (Chile)
suggested  a  high  orbital  eccentricity and the presence of pulsations
causing the measured radial velocity curve to deviate significantly from that expected 
from Keplerian motion \citep{Lorenzo2014}.
Since, according to the model, supergiant magnetic fields can 
play a major role for the generation of outbursts, we carried out the first search ever
for magnetic fields in the two brightest SFXTs accessible with the VLT. In the following, we
present the results of our 
magnetic field measurements and discuss the detected spectral variability.

\section{Observations and magnetic field measurements}
\label{sect:obs}

\begin{figure}
\centering
\includegraphics[width=0.48\textwidth]{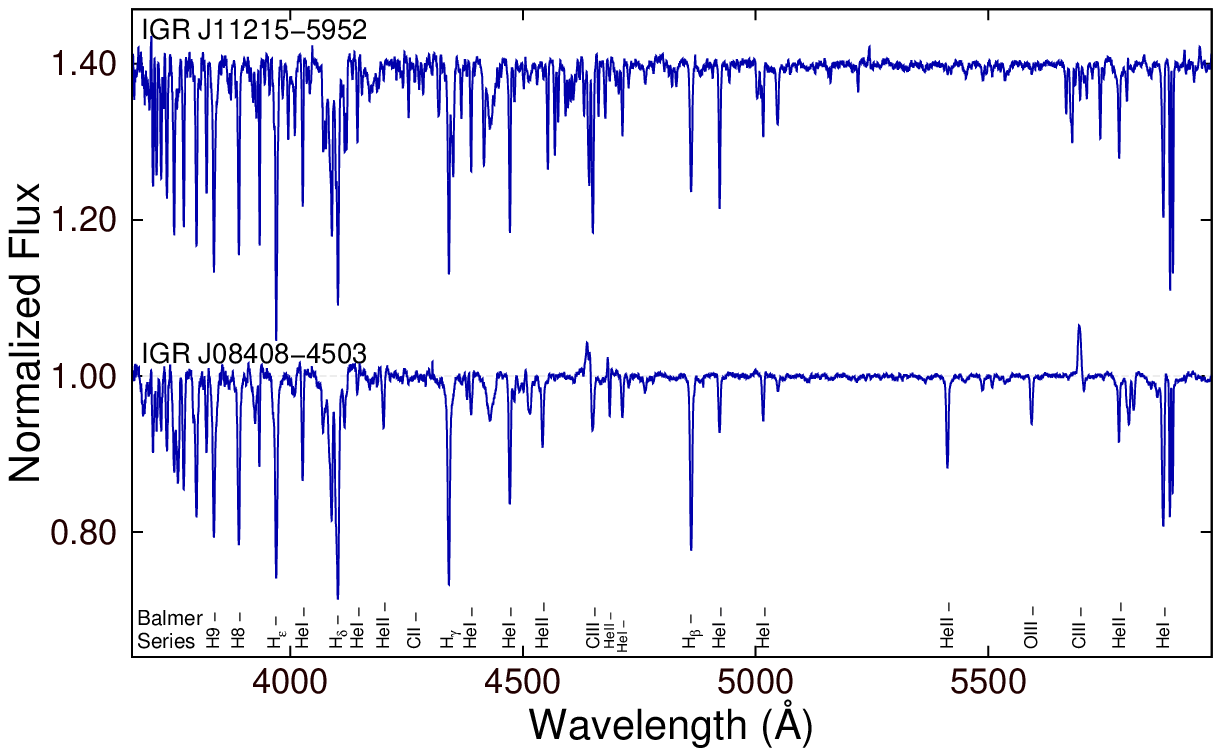}
\caption{ 
Normalised FORS\,2 Stokes~$I$
spectra of IGR\,J08408$-$4503 and IGR\,J11215$-$5952.
Well known spectral lines are indicated at the bottom.
The spectrum of IGR\,J11215$-$5952 was vertically offset by 0.4 for clarity.
}
\label{fig:norm}
\end{figure}

FORS\,2 spectropolarimetric observations of both systems were obtained
from 2016 May~14 to 2017 February~9.
The FORS\,2 multi-mode instrument is equipped with polarisation analysing optics
comprising super-achromatic half-wave and quarter-wave phase retarder plates,
and a Wollaston prism with a beam divergence of 22$\arcsec$ in standard
resolution mode. 
We used the GRISM 600B and the narrowest available slit width
of 0$\farcs$4 to obtain a spectral resolving power of $R\sim2000$.
The observed spectral range from 3250 to 6215\,\AA{} includes all Balmer lines,
apart from H$\alpha$, and numerous helium lines.
For the observations, we used a non-standard readout mode with low 
gain (200kHz,1$\times$1,low), which provides a broader dynamic range, hence 
allowed us to reach a higher signal-to-noise ratio (SNR) in the individual spectra.
The  spectral  appearance  of IGR\,J08408$-$4503 and IGR\,J11215$-$5952 in the FORS\,2 spectra
is presented in Fig.~\ref{fig:norm}.
A description of the assessment of the presence of a longitudinal magnetic field
using FORS\,1/2 spectropolarimetric observations was presented 
in our previous work (e.g.\ \citealt{Hubrig2004a,Hubrig2004b}, 
and references therein).
Rectification of the $V/I$ spectra was
performed in the way described by \citet{Hubrig2014}.
Null profiles, $N$, are calculated as pairwise differences from all available 
$V$ profiles so that the real polarisation signal should cancel out.  
From these, 3$\sigma$-outliers are identified and used to clip 
the $V$ profiles.  This removes spurious signals, which mostly come from cosmic
rays, and also reduces the noise. A full description of the updated data 
reduction and analysis will be presented in a separate paper (Sch\"oller et 
al., in preparation, see also \citealt{Hubrig2014}).
The mean longitudinal magnetic field, $\left< B_{\rm z}\right>$, is 
measured on the rectified and clipped spectra based on the relation 
following the method suggested by \citet{Angel1970}
\begin{eqnarray} 
\frac{V}{I} = -\frac{g_{\rm eff}\, e \,\lambda^2}{4\pi\,m_{\rm e}\,c^2}\,
\frac{1}{I}\,\frac{{\rm d}I}{{\rm d}\lambda} \left<B_{\rm z}\right>\, ,
\label{eqn:vi}
\end{eqnarray} 

\noindent 
where $V$ is the Stokes parameter that measures the circular polarization, $I$
is the intensity in the unpolarized spectrum, $g_{\rm eff}$ is the effective
Land\'e factor, $e$ is the electron charge, $\lambda$ is the wavelength,
$m_{\rm e}$ is the electron mass, $c$ is the speed of light, 
${{\rm d}I/{\rm d}\lambda}$ is the wavelength derivative of Stokes~$I$, and 
$\left<B_{\rm z}\right>$ is the mean longitudinal (line-of-sight) magnetic field.

The longitudinal magnetic field was measured in two ways: using the entire spectrum
including all available lines, among them 13 strong He lines,
or using exclusively hydrogen lines, 13 in total.
Furthermore, we have carried out Monte Carlo bootstrapping tests. 
These are most often applied with the purpose of deriving robust estimates of standard errors
(e.g.\ \citealt{Steffen2014}). 
The measurement uncertainties obtained before and after the Monte Carlo bootstrapping tests were found to be 
in close agreement, indicating the absence of reduction flaws. 
The results of our magnetic field measurements, those for the entire spectrum
or only the hydrogen lines are presented in 
Table~\ref{tab:log_meas}. The orbital phases of IGR\,J08408$-$4503 were calculated relative
to a zero phase corresponding to the periastron passage at
$T_{\rm per}= 2454654.04$ \citep{Gamen2015}. 
The same zero phase as in the work of \citet{Lorenzo2014} corresponding to the date of
the first FEROS observation was adopted for IGR\,J11215$-$5952.
The X-ray outbursts occurred at orbital phases close to 0.4, following \citet{Lorenzo2014}.  

\begin{figure}
\centering
\includegraphics[width=0.23\textwidth]{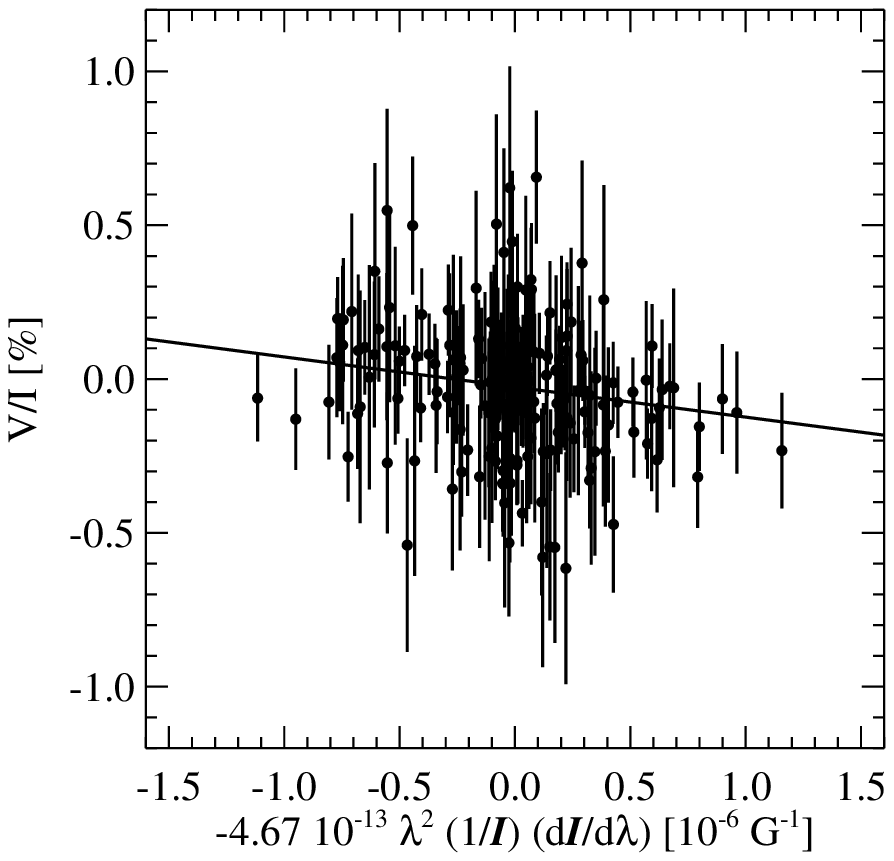}
\includegraphics[width=0.23\textwidth]{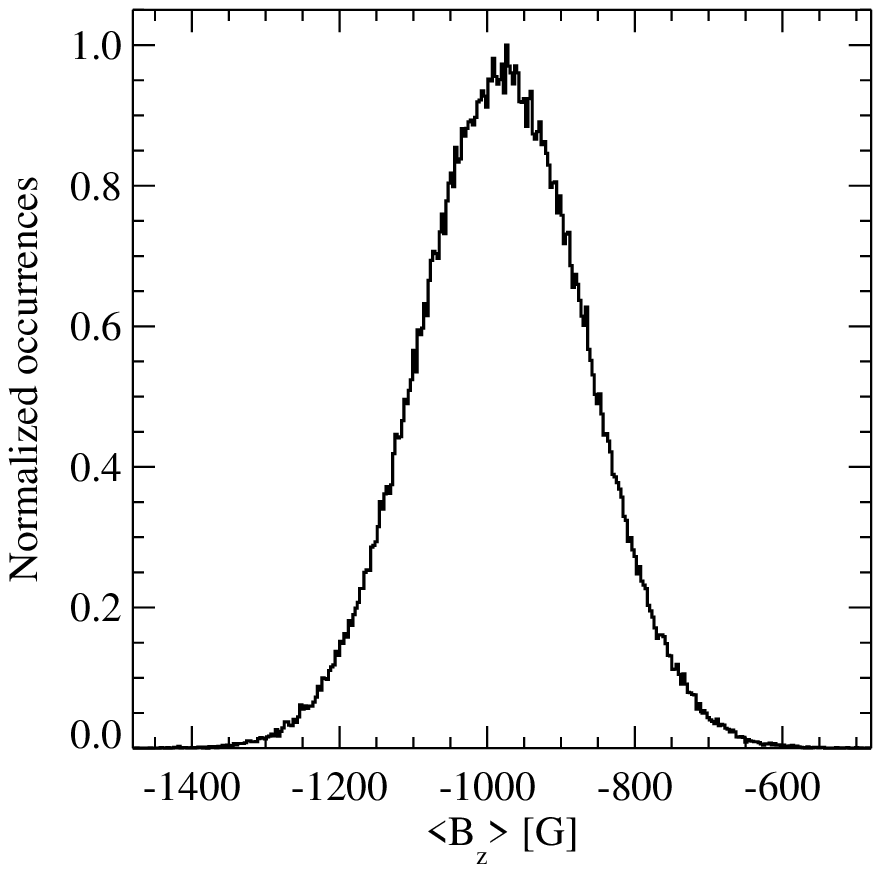}
\caption{
{\it Left panel}: Linear fit to Stokes~$V$ spectrum obtained for the FORS\,2 
observation of IGR\,J11215$-$5952 on MJD\,57522.0581.
{\it Right panel}: Distribution of the longitudinal magnetic field values $P(\left<B_{\rm z}\right>)$, 
which were obtained via bootstrapping.
From this distribution follows the most likely value for the longitudinal 
magnetic field $\left< B_{\rm z}\right>_{\rm hydr}=-978\pm308$\,G.
}
\label{fig:igr_1}
\end{figure}

\begin{figure}
\centering
\includegraphics[width=0.23\textwidth]{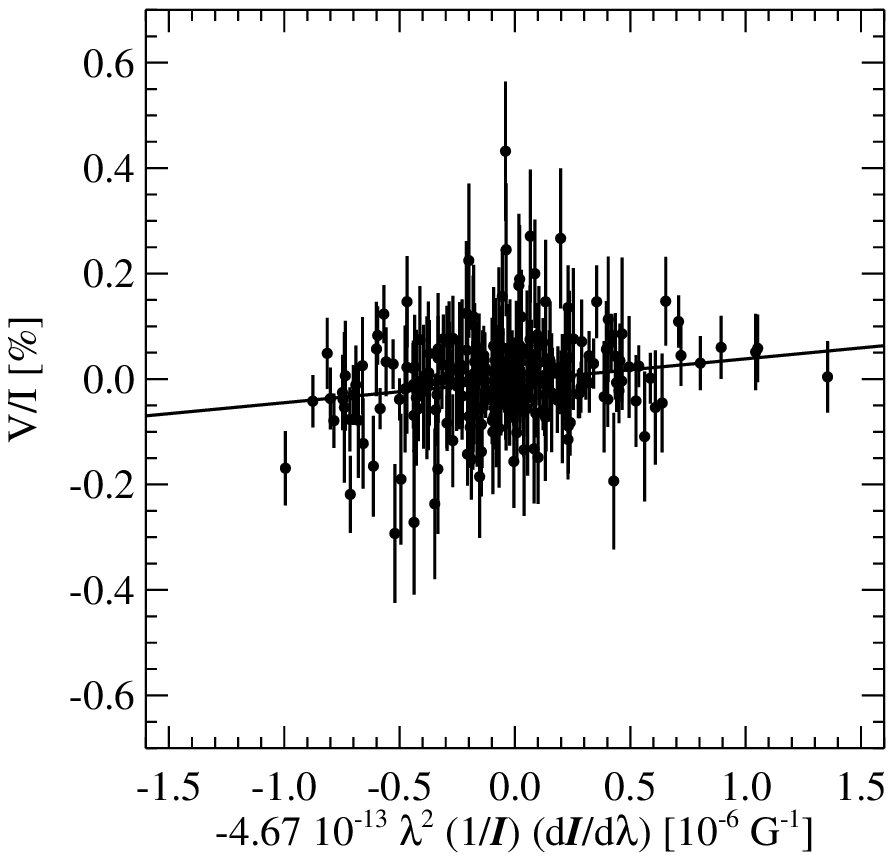}
\includegraphics[width=0.23\textwidth]{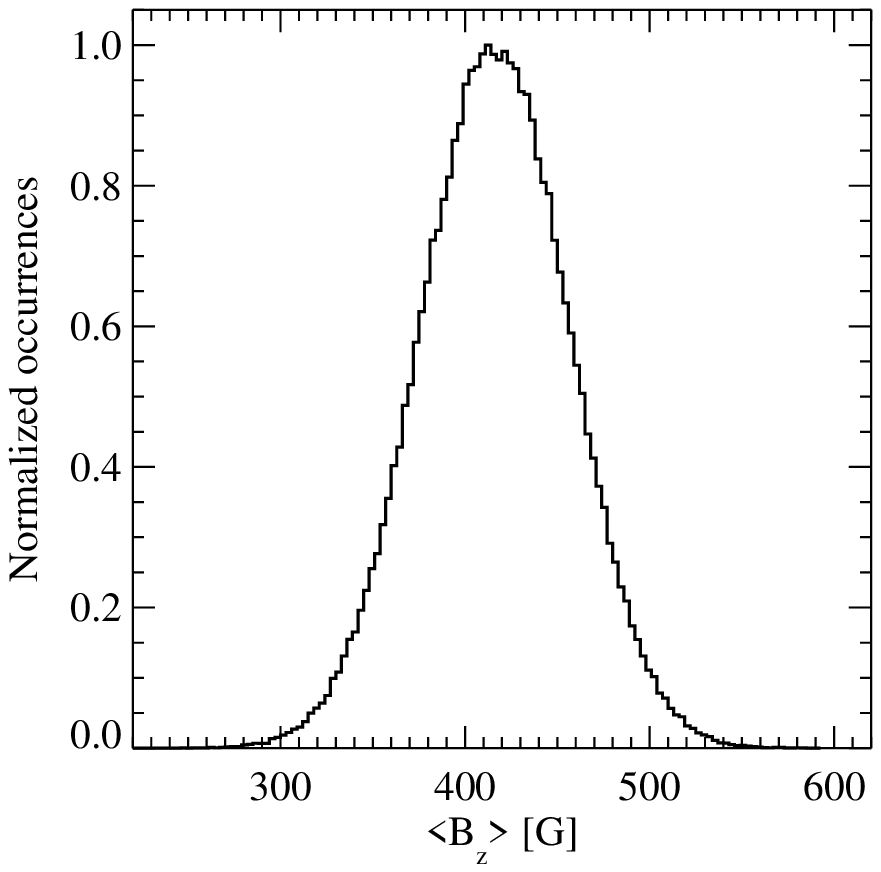}
\caption{
{\it Left panel}: Linear fit to Stokes~$V$ spectrum obtained for the FORS\,2 
observation of IGR\,J11215$-$5952 on MJD\,57738.2863.
{\it Right panel}: Distribution of the longitudinal magnetic field values $P(\left<B_{\rm z}\right>)$, 
which were obtained 
via bootstrapping. From this distribution follows the most likely value for the longitudinal 
magnetic field $\left< B_{\rm z}\right>_{\rm hydr}=416\pm110$\,G.
}
\label{fig:J11_1}
\end{figure}

\begin{figure}
\centering
\includegraphics[width=0.48\textwidth]{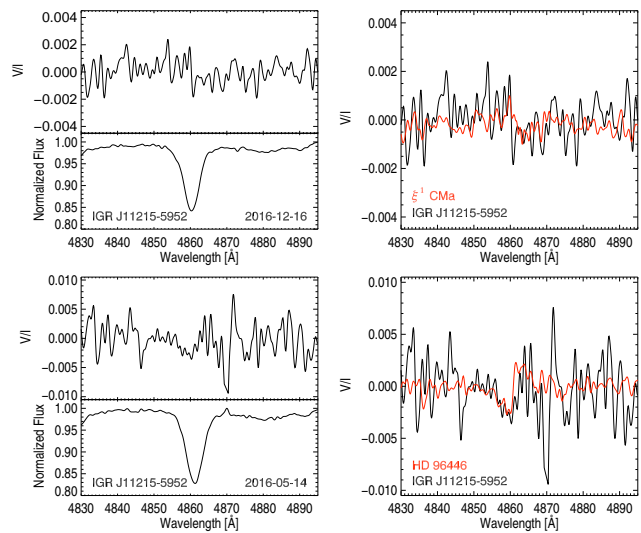}
\caption{ 
{\it Left panel}: Stokes~$V$  and Stokes~$I$ spectra of IGR\,J11215$-$5952 in the spectral region around the H$\beta$ line
at two different epochs.
{\it Right panel}: Stokes~$V$ spectra of IGR\,J11215$-$5952 are overplotted with the Stokes $V$ spectra
of the two well-known magnetic early-B type stars  $\xi^1$\,CMa ($\left< B_{\rm z}\right>_{\rm hydr}=360\pm49$\,G)
and HD\,96446 ($\left< B_{\rm z}\right>_{\rm hydr}=-1590\pm74$\,G) for best visibility of the Zeeman features.
}
\label{fig:zf}
\end{figure}

The magnetic field of IGR\,J08408$-$4503 was measured on four 
nights in 2016, but no detection at a significance level of 3$\sigma$
was achieved in any of the measurements. The magnetic field, if present,
would likely be rather weak or variable. The highest value for the longitudinal
magnetic field, $\left< B_{\rm z}\right>_{\rm hydr}=-184\pm97$\,G at a significance level
of 1.9$\sigma$ was measured in the observation obtained on 2016 December 14.
For IGR\,J11215$-$5952, we obtain 3.2$\sigma$ ($\left< B_{\rm z}\right>_{\rm hydr}=-978\pm308$\,G)
and 3.8$\sigma$ ($\left< B_{\rm z}\right>_{\rm hydr}=416\pm110$\,G) detections on two observing nights, 
on 2016 May 14 and 2016 December 16,
respectively. In Figs~\ref{fig:igr_1} and \ref{fig:J11_1}, we show linear regressions in the plots 
$V/I$ against $-4.67\,10^{-13} \lambda^2 (1/I) ({\rm d}I/{\rm d}\lambda)$ together with 
the results of the Monte Carlo bootstrapping tests. No detection was achieved in the diagnostic $N$ spectra,
indicating the absence of spurious polarization signatures.
The slopes of the lines fitted to the data directly translate into the values for $\left<B_{\rm z}\right>$.
In Fig.~\ref{fig:zf}, we present Stokes $V$ spectra of IGR\,J11215$-$5952 obtained on these two nights 
in the spectral region around the H$\beta$ line. For best visibility of the Zeeman features, we overplot 
the Stokes~$V$ spectra of IGR\,J11215$-$5952 with the Stokes~$V$ spectra
of the two well-known magnetic early B-type stars HD\,96446 and $\xi^1$\,CMa.

Regarding the significance of the magnetic field detections in massive stars at 
significance levels around 3$\sigma$, we note that the two clearly magnetic Of?p stars
HD\,148937 and CPD $-$28$^\circ$ 5104 have been
for the first time detected as magnetic in our FORS\,2 observations at significance 
levels of 3.1$\sigma$ and 3.2$\sigma$, 
respectively (\citealt{Hubrig2008,Hubrig2011}).
The detection of the magnetic field in IGR\,J11215$-$5952 at significance levels of 3.2$\sigma$ and
3.8$\sigma$ indicates that this target likely possesses a kG magnetic field.

According to the orbital phases presented by \citet{Lorenzo2014}, the negative magnetic field
$\left< B_{\rm z}\right>_{\rm hydr}=-978\pm308$\,G is detected at the orbital phase 0.958 while 
the positive magnetic field $\left< B_{\rm z}\right>_{\rm hydr}=416\pm110$\,G is detected at the phase 0.272.
The phase 0 in \citet{Lorenzo2014} corresponds to their first optical observation performed on MJD\,54072.33.
and has nothing to do with the true periastron passage.
Using for IGR\,J11215$-$5952 the estimate of the stellar radius $R= 40\pm 5\,R_\odot$ 
and $v \sin i = 50$\,km\,s$^{-1}$ \citep{Lorenzo2014}, 
the rotation period $P_{\rm rot}$ is expected to be equal or less than 40.5\,d, i.e. its length 
can be at least four times shorter than the orbital period of 164.6\,d. 
We can speculate that if the ratio between  $P_{\rm rot}$
and $P_{\rm orb}$ is indeed a factor of four, then the same region of the stellar surface of 
the optical counterpart of IGR\,J11215$-$5952, HD\,306414,
would face the compact component during the periastron passage.

\section{Spectral variability}
\label{sect:var}

\begin{figure}
\centering
\includegraphics[width=0.48\textwidth]{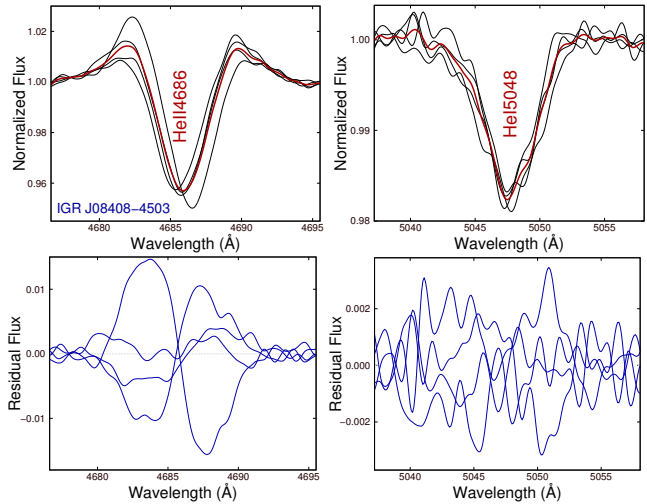}
\caption{ 
The variability of the \ion{He}{ii}~4686 line (left side) in the spectra of IGR\,J08408$-$4503
on four different orbital phases.  The upper and lower panels show the overplotted profiles 
and the differences between the individual and the average (red) line profiles.
No significant variability exceeding the noise level is detected in the \ion{He}{i}~5048 line (right side)
}
\label{fig:var1}
\end{figure}

\begin{figure}
\centering
\includegraphics[width=0.48\textwidth]{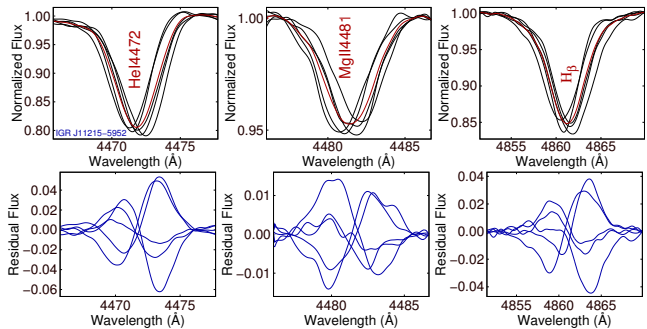}
\caption{ 
The variability of the spectral lines \ion{He}{i}~4472, H$\beta$, and
\ion{He}{i}~5048 in IGR\,J11215$-$5952 on five different observing nights.
The upper and lower panels 
show the overplotted profiles and the differences between the individual and the average (red) line profiles.
}
\label{fig:var2}
\end{figure}

Since the presence of 
pulsations is frequently reported in early type supergiants (see the most recent literature overview 
presented in the work of \citealt{Lorenzo2014}), 
we checked  the stability of the line profiles in the FORS\,2 spectra of IGR\,J08408$-$4503 and 
IGR\,J11215$-$5952 between the nights   
and along the full sequences of sub-exposures during each observing night. 
While only the \ion{He}{ii}~4686 line was detected as variable in the four FORS\,2 spectra of IGR\,J08408$-$4503
distributed over 20 days
(Fig.~\ref{fig:var1}), most line profiles in the five spectra of IGR\,J11215$-$5952 distributed  
over 271 days show intensity variations and small 
radial velocity shifts (Fig.~\ref{fig:var2}). As discussed by \citet{Lorenzo2014}, the spectral variability of 
IGR\,J11215$-$5952 can be caused by orbital movement combined with pulsational variability.
The detected spectral variability can, however, also be caused by a surface inhomogeneous
distribution of several elements as is usually observed in magnetic early-type Bp stars. Future careful
high-resolution spectroscopic monitoring of this target would be useful to 
identify the origin of the detected variability.

\begin{figure}
\centering
\includegraphics[width=0.45\textwidth]{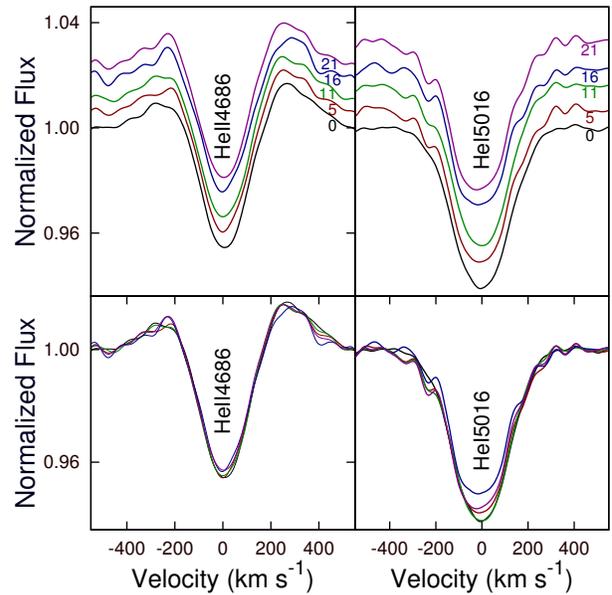}
\caption{ 
The behaviour of \ion{He}{ii}~4686 and \ion{He}{i}~5016 in the spectra of IGR\,J08408$-$4503
obtained for individual subexposures belonging to the observation on 2016 December 5.
In the upper row, we present the line profiles shifted in vertical direction for best visibility.
The time difference (in minutes) between each subexposure and the start of
observations is given close to each profile. The lower row shows all profiles
overplotted.
}
\label{fig:var3}
\end{figure}

To search for variability on short time scales, we have compared for each observation
the Stokes~$I$ line profiles recorded in each subexposure.
Only very low intensity variability is detected in IGR\,J08408$-$4503 in the \ion{He}{ii}~4686 line, while 
it is stronger for the \ion{He}{i} lines. In Fig.~\ref{fig:var3}, we present the individual Stokes 
$I$ profiles of \ion{He}{ii}~4686  and \ion{He}{i}~5015.
The line profiles in the FORS\,2 spectra of IGR\,J11215$-$5952 look identical within the noise.

\section{Discussion}
\label{sect:disc}

The formation, evolution, and fate of  SFXTs is only partly understood, due to
our limited knowledge about the evolution of massive stars.
Obviously, understanding these systems is of immense interest, 
as they probably evolve to NS-NS or NS-BH binaries and are progenitors of gravitational wave signals.

The persistent presence of H$\alpha$ emission in IGR\,J08408$-$4503 can be explained by different 
physical processes, such as a departure from local thermodynamic equilibrium or/and by light scattering in a geometrically 
extended atmosphere.
On the other hand, it is also possible that
the star supports a centrifugal magnetosphere with 
magnetically confined material forming a disk-like structure around the magnetic equatorial plane.
Depending on the dipole magnetic field strength and stellar rotation rate, such a magnetosphere
would extend to the Alfv\'en radius of several stellar radii, allowing magnetized clumps of stellar wind
to trigger sporadic reconnection during the periastron passage
and mass accretion onto the compact companion, which would then result in an X-ray flare.
Magnetized clumps
caught in the region between the corotating Keplerian radius and the Alfv\'en radius
would be centrifugally supported against infall, and so build up to a much denser centrifugal 
magnetosphere.
The non-detection of a magnetic field in our observations of IGR\,J08408$-$4503,
which are all in orbital phases close to the periastron passage,
might be explained by a magnetic field configuration in which we view the magnetic equator of the dipole
during these phases.
This would also require that the supergiant either rotates synchronously or in resonance with the orbit.

Our spectropolarimetric observations of IGR\,J11215$-$5952 
revealed the presence of a magnetic field on two occasions.
This target is the only SFXT where strictly periodic X-ray outbursts have been observed, 
repeating every 164.6\,d (\citealt{Sidoli2006,Sidoli2007,Romano2009}).
To explain these short periodic outbursts, \citet{Sidoli2007} proposed that they are triggered by the passage 
of the neutron star inside an equatorial enhancement of the outflowing supergiant wind, focussed on a plane 
inclined with respect to the orbit. 
This  configuration of the line-drive stellar wind might be magnetically channeled \citep{ud-Doula2002}.
The effectiveness of the stellar magnetic field in focussing the wind is indicated by the
wind magnetic confinement parameter $\eta$
defined as $\eta= B_{*} ^2 R_{*}^2$ / ($\dot M_{\rm w}\upsilon_{\infty}$),
where B$_{*}$ is the strength of the magnetic field at the surface of the supergiant, R$_{*}$ is the stellar radius
(R$_{*}$=40~R$_\odot$), $\upsilon_{\infty}$ is the wind terminal velocity ($\upsilon_{\infty}=1200~km~s^{-1}$)
and $\dot{M}_{\rm w}$ is the wind mass loss rate ($\dot M_{\rm w}=10^{-6}$\,M$_\odot$\,yr$^{-1}$; \citealt{Lorenzo2014}).
Adopting B$_{*} \ge$ 0.7\,kG at the magnetic equator, we estimate $\eta \ge$ 500, 
implying a wind 
confinement, up to an Alfv\'en radius $R_A=\eta^{1/4}R_{*} \ge 4.73R_{*}$  \citep{ud-Doula2002}. This 
radial distance is 
compatible with the orbital separation at periastron in IGR\,J11215$-$5952,
where the orbital eccentricity is high (e$>$0.8; \citealt{Lorenzo2014}). 
The measured magnetic field strength in IGR\,J11215$-$5952 reported here for the first time is high enough to channel 
the stellar wind on the magnetic equator, supporting the scenario proposed by \citet{Sidoli2007} to explain the 
short periodic outbursts in this SFXT. 

Long-term intensity and radial velocity variability of the wind line \ion{He}{ii}~4686 and short-term 
intensity variability in the \ion{He}{i} lines were detected in IGR\,J08408$-$4503 and can probably be explained
by the strong wind in the hot supergiant and by obscuration caused by the surrounding material.
The long-term variability intensity and radial velocity variability of the hydrogen and He lines in IGR\,J11215$-$5952 
was already detected by \citet{Lorenzo2014} and is probably produced by stellar pulsations or surface
chemical spots.

Because of the faintness of SFXTs -- most of them have a visual magnitude $m_V \ge 12$, up to  $m_V \ge 31$ 
(\citealt{Sidoli2017,Persi2015}) --
no high-resolution spectropolarimetric observations were carried out for them so far and the presented FORS\,2 
observations are the first to explore the magnetic nature of the optical counterparts.
Future spectropolarimetric observations 
of a representative sample of SFXTs are urgently needed to be able to draw solid conclusions about
the role of magnetic fields in the wind accretion process and to 
constrain the conditions that enable the presence of magnetic
fields in massive binary systems.

\section*{Acknowledgments}
\label{sect:ackn}
Based on observations obtained in the framework of the ESO Prgs.\ 097.D-0233(A) and 098.D-0185(A).
LS acknowledges financial contribution from the grant from PRIN-INAF 2014 "Toward a unified picture of accretion in High Mass
X-Ray Binaries". AK acknowledges financial support from RFBR grant 16-02-00604A. We thank the referee Jerzy Madej
for his useful comments.

\label{lastpage}


\begin{thebibliography}{}

\bibitem[Angel \& Landstreet(1970)]{Angel1970}
Angel J.~R.~P., Landstreet J.~D., 1970,
ApJ, 160, L147

\bibitem[Appenzeller et al.(1998)]{Appenzeller1998} 
%
Appenzeller I., et al., 1998,
The ESO Messenger, 94, 1

\bibitem[Cowperthwaite et al.(2017)]{Cowperthwaite2017}
Cowperthwaite P.~S., et al., 2017,
submitted to ApJ, {\sl also:} arXiv:1710.05840

\bibitem[Gamen et al.(2015)]{Gamen2015}
Gamen R., et al., 2015,
A\&A, 583, L4

\bibitem[Hubrig et al.(2004a)]{Hubrig2004a}
Hubrig S., Kurtz D.~W., Bagnulo S., Szeifert T., Sch{\"o}ller M., Mathys G., Dziembowski W.~A., 2004a,
A\&A, 415, 661

\bibitem[Hubrig et al.(2004b)]{Hubrig2004b}
Hubrig S., Szeifert T., Sch{\"o}ller M., Mathys G., Kurtz D.~W., 2004b,
A\&A, 415, 685

\bibitem[Hubrig, Sch\"oller, \& Kholtygin(2014)]{Hubrig2014}
Hubrig S., Sch\"oller M., Kholtygin A.~F., 2014,
MNRAS, 440, L6

\bibitem[Hubrig et al.(2008)]{Hubrig2008}
Hubrig S., Sch\"oller M., Schnerr R.~S., Gonz\'alez J.~F., Ignace R., Henrichs H., 2008,
A\&A, 490, 793

\bibitem[Hubrig et al.(2011)]{Hubrig2011}
Hubrig S., et al., 2011,
A\&A, 528, A151

\bibitem[Liu et al.(2006)]{Liu2006}
Liu Q.~Z., van Paradijs J., and van den Heuvel E.~P.~J., 2006, A\&A, 455, 1165

\bibitem[Lorenzo et al.(2014)]{Lorenzo2014}
Lorenzo J., Negueruela I., Castro N., Norton A.~J., Vilardell F., Herrero A., 2014,
A\&A, 562, A18

\bibitem[Martinez-Nunez et al.(2017)]{Martinez2017}
Martinez-Nunez S., et al., 2017,
SSRv, 212, 59

\bibitem[Persi et al.(2015)]{Persi2015}
Persi P., Fiocchi M., Tapia M., Roth M., Bazzano A., Ubertini P., Parisi, P.,\ 2015, AJ, 150, 21
%
%

\bibitem[Romano et al.(2009)]{Romano2009}
Romano P., Sidoli L., Cusumano G., Vercellone S., Mangano V., Krimm H.~A., 2009,
ApJ, 696, 2068

\bibitem[Shakura et al.(2014)]{Shakura2014}
Shakura N., Postnov K., Kochetkova A., Hjalmarsdotter L., 2012,
MNRAS, 442, 2325

\bibitem[Shakura et al.(2012)]{Shakura2012}
Shakura N., Postnov K., Sidoli L., Paizis A., 2014,
MNRAS, 420, 216

\bibitem[Shakura et al.(2015)]{Shakura2015}
Shakura N.~I., Postnov K.~A., Kochetkova A.~Yu., Hjalmarsdotter L., Sidoli L., Paizis A., 2015.
ARep, 59, 645

\bibitem[Shakura \& Postnov(2017)]{Shakura2017}
Shakura N., Postnov K., 2017,
to appear in PoS Accretion Processes in Cosmic Sources, September 5-10, 2016, St-Petersburg,
{\sl also:} arXiv:1702.03393

\bibitem[Sidoli et al.(2006)]{Sidoli2006}
Sidoli L.,  Paizis A., Mereghetti S., 2006, 
A\&A, 450, L9

\bibitem[Sidoli et al.(2007)]{Sidoli2007}
Sidoli L.,  Romano P., Mereghetti S., Paizis A, Vercellone S., Mangano V., G{\"o}tz D., 2007, 
A\&A, 476, 1307

\bibitem[Sidoli(2017)]{Sidoli2017}
Sidoli L., 2017,
submitted to the Proceedings of the ``XII Multifrequency Behaviour of High Energy Cosmic Sources Workshop'',
{\sl also:} arXiv:1710.03943,

\bibitem[Sota et al.(2014)]{Sota2014}
Sota A., et al., 2014,
ApJS, 211, 10

\bibitem[Steffen et al.(2014)]{Steffen2014}
Steffen M., Hubrig S., Todt H., Sch\"oller M., Hamann W.-R., Sandin C., Sch\"onberner, D.,\ 2014, A\&A, 570, A88

\bibitem[van den Heuvel \& Heise(1972)]{Heuvel1972}
van den Heuvel E.~P.~J., Heise, J., 1972,
Nature Physical Science, 239, 67

\bibitem[ud-Doula \& Owocki(2002)]{ud-Doula2002}
ud-Doula A., and Owocki S.~P., 2002, ApJ 576, 413

\end{thebibliography}
\end{document}